\documentclass[
reprint,
groupedaddress,
longbibliography,
amsmath,amssymb,
aps,
prl,
floatfix,
]{revtex4-2}

\usepackage{graphicx}
\usepackage{dcolumn}
\usepackage{bm}
\usepackage{physics}

\usepackage{graphicx}%
\usepackage{xcolor}
\usepackage{cancel}
\usepackage{dcolumn}%
\usepackage{bm}%
\usepackage[colorlinks, linkcolor=blue, citecolor=blue, urlcolor=blue]{hyperref}%
\usepackage{booktabs}

\usepackage{subfigure} 

\newcommand{\quadh}{ \Delta\mathcal{H}}

\begin{document}

\preprint{APS/123-QED}

\author{Preethi Basani}
\author{Varsha Subramanyan}
\author{Smitha Vishveshwara}
\affiliation{Department of Physics, University of Illinois at Urbana-Champaign, Urbana, IL, USA}

\title{Symmetries and dynamics of quantum Hall bulk anyons in quadratic potentials}


\begin{abstract}
We study two-particle coherent states and their dynamics in the lowest Landau level (LLL) under the influence of quadratic potentials. We focus on generalized coherent states that describe Abelian anyons in the LLL and are associated with the $\mathfrak{su}(1,1)$ Lie algebra. We draw on parallels with quantum optics and symmetry properties of the coherent states considered here to analytically calculate quantities such as a bunching parameter, which depends on quantum statistics, as well as coherent state trajectories under the influence of generic quadratic potentials. Our results show that in unbounded saddle potentials, the bunching parameter governs the trajectories which show exponentially diverging behavior in a manner that depends on quantum statistics.  In bounded elliptical potentials, the bunching parameter is oscillatory and its maximum magnitude depends on the eccentricity of the applied potential. We draw connections between our analyses and the key concepts that underlie anyon detection in recent experiments.

\end{abstract}
\maketitle

\begin{figure}[h!]
  \centering
  \subfigure[]{%
    \includegraphics[width=0.48\linewidth]{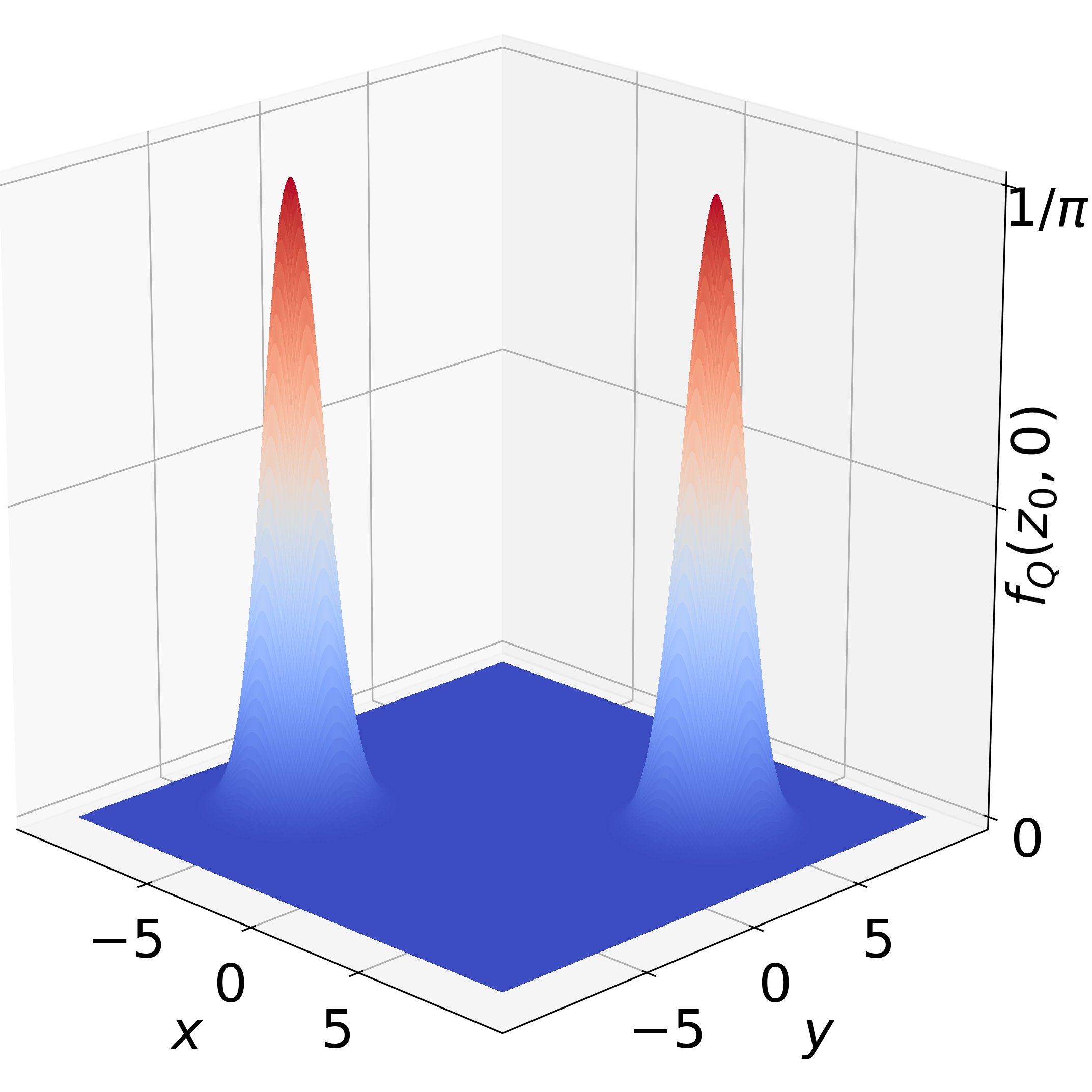}
    \label{fig:initial_anyon_state}%
  }
    \hfill
  \subfigure[]{%
    \includegraphics[width=0.48\linewidth]{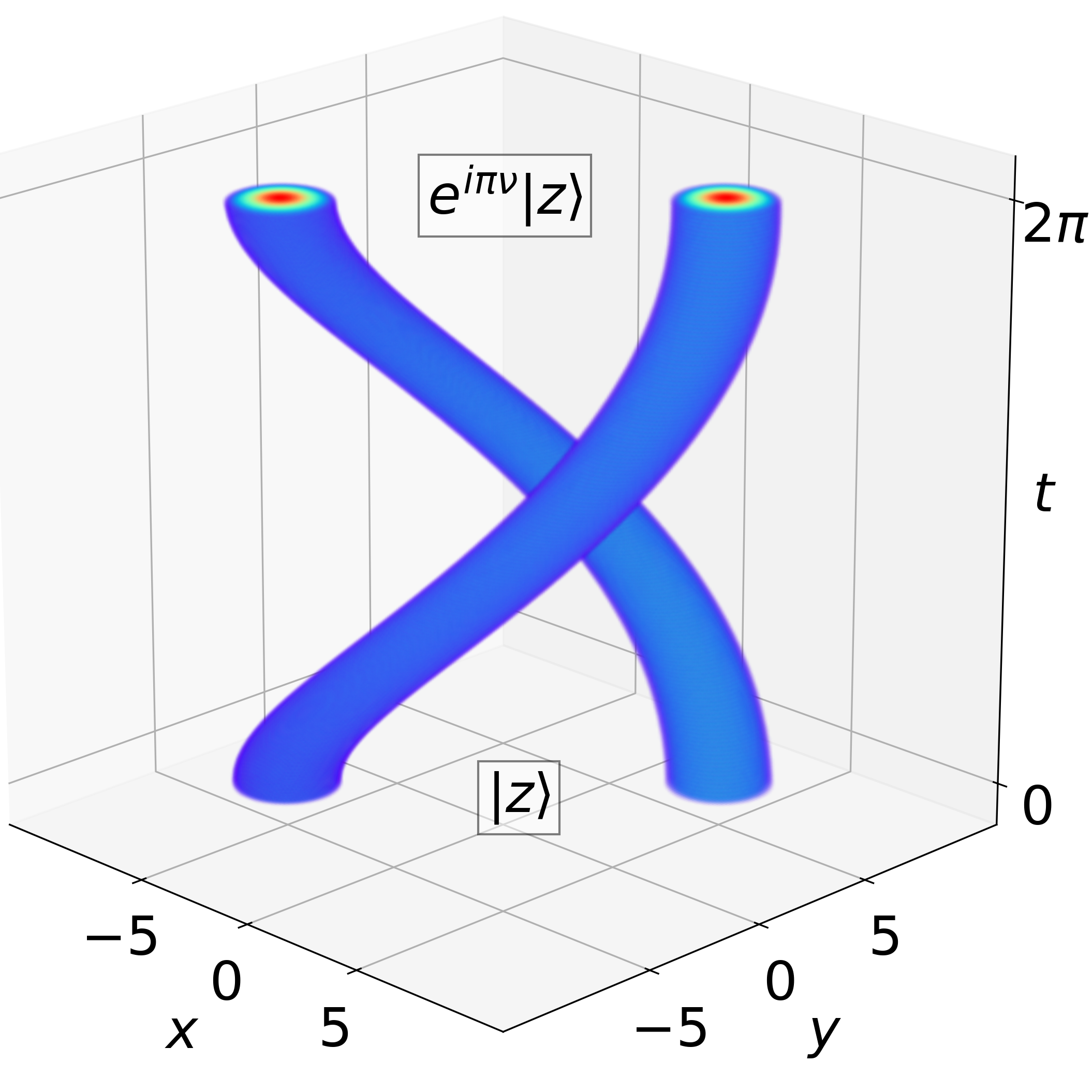}
    \label{fig:worldlines}%
  }
  \caption{(a) Spatial plot of two anyons at time $t = 0$ that have their center-of-mass at the origin, as characterized by the Husimi-Q distribution $f_Q(z_0,0)$ derived in the text.
  (b) Time evolution of the distribution under the influence of a harmonic potential of strength $U$(shown as contour plots for each time step). Spatial coordinates are in units of the magnetic length $l_B$ and time $t$ in units of $U l_B^2/\hbar$. The trajectories act as worldline proxies, showing explicit exchange and acquisition of the fractional phase $e^{i\pi\nu}$, where $0<\nu<1$.}
  \label{fig:initial_anyon_state_plus_worldlines}
\end{figure}

Anyons--identical particles that obey fractional statistics--have enjoyed age-old appeal in the context of the two-dimensional fractional quantum Hall (FQH) system as a prospective arena for their realization\cite{leinaasTheoryIdenticalParticles1977a,wilczekQuantumMechanicsFractionalSpin1982,laughlinAnomalousQuantumHall1983,arovasFractionalStatisticsQuantum1984,goldinDiffeomorphismGroupsGauge1983,prangeQuantumHallEffect1990,comtetTopologicalAspectsLowdimensional1999,jainCompositeFermions2007}. Interpolating between bosons and fermions, Abelian anyons pick up a complex phase $e^{i\pi\nu}$ on exchange, where $\nu$ is a fraction between $0$ and $1$(depicted in  Fig. \ref{fig:worldlines}) .  In recent years, the experimental detection of FQH quasiparticle excitations that have anyonic signatures in beamsplitter and interferometric settings has caused a resurgence in attention \cite{bartolomeiFractionalStatisticsAnyon2020,nakamuraDirectObservationAnyonic2020,willettInterferenceMeasurementsNonAbelian2023,leePartitioningDilutedAnyons2023,kunduAnyonicInterferenceBraiding2023,samuelsonSlowQuasiparticleDynamics2025}. 
It has revitalized decades-long studies of connecting theoretical and mathematical underpinnings of FQH states, fabrication and materials issues in physical systems, innovative experimental geometries, and implications for topological quantum computation (which requires a generalization to non-Abelian anyons) \cite{nayakNonAbelianAnyonsTopological2008}. While most of the treatments have focused on gapless edge-state FQH particles, which are amenable to transport and manipulation, the excitations are ultimately rooted in the physics of their gapped parent quantum Hall(QH) bulk. Here, we therefore revisit QH bulk anyons in the lowest Landau level(LLL)\cite{hanssonDimensionalReductionAnyon1992,kjonsbergAnyonDescriptionLaughlin1997a,senQuasiparticlePropagationQuantum2008,paredes$frac12$AnyonsSmallAtomic2001,ben-shachDetectingNonAbelianAnyons2013a,schineSyntheticLandauLevels2016,papicImagingAnyonsScanning2018} and develop a theoretical framework to study their behavior in the presence of quadratic potentials, which are ubiquitous in QH systems. 
\newline

The framework we present provides a comprehensive description of generalized coherent states in the LLL under the influence of quadratic potentials of both the bounded and unbounded variety and their associated dynamics. We build on a beautiful description of localized Abelian anyons in Laughlin FQH fluids involving such generalized coherent states constructed through fractional angular momentum states of the LLL that satisfy anyon boundary conditions.
The description benefits from the symmetry properties of quantum states in the LLL, which are invariant under area-preserving transformations generated by the $\mathfrak{sl}(2,\mathbb{R})$ Lie algebra (or isomorphically, $\mathfrak{su}(1,1)$).   Since quadratic operators in the LLL are also generated by $\mathfrak{su}(1,1)$, our method provides an analytically tractable method of extracting such dynamics. By abstracting the many-body complexity of anyon excitations to two-particle dynamics, our method focuses on fundamental aspects that arise purely due to the exchange statistics of fractional excitations.

In what follows, we establish the coherent states description for a pair of anyons and the symmetry-based framework for treating quadratic potentials. We demonstrate how invoking parallels with quantum optics offers the desired analytic tools. We perform an in-depth study of dynamics in the presence of unbounded potentials, which are relevant to beam-splitter, anyon scattering, and Hanbury Brown-Twiss physics\cite{brownCorrelationPhotonsTwo1956,hongMeasurementSubpicosecondTime1987,baymPhysicsHanburyBrownTwiss1998}, and of bounded potentials, which are relevant to interferometry in the LLL\cite{feldmanRobustnessQuantumHall2022}.


\par \textit{Coherent state formalism.}- 
Coherent states offer a well-established starting point for describing localized excitations in the QH bulk. In particular, in the LLL, they form symmetrically distributed wave-packets in two-dimensional space that respect the minimum uncertainty condition set by the magnetic length in all directions. 
\par 
In describing Abelian anyons using coherent states, the key is to demand that the states obey fractional exchange statistics. In the simplest case of two particles with positions $\vec{r}_1 = (x_1,y_1)$ and $\vec{r}_2 = (x_2,y_2)$, this can be achieved with a coordinate transform to center-of-mass (COM) coordinates $\vec{R} = \frac{1}{2}\left(\vec{r}_1+ \vec{r}_2\right) = (X,Y)$ and relative coordinates $ \vec{r} = \vec{r}_1 - \vec{r}_2 = (x,y)$. In these coordinates, the state is symmetric under sign change in the COM coordinate and the state picks up a fractional phase under sign change of the relative coordinate, giving the boundary conditions:
\begin{equation}\label{eq:boundary-conditions}
    \psi_{\text{COM}}(-\vec{R}) = \psi_{\text{COM}}(\vec{R}), \, \psi_{\text{rel}}(-\vec{r}) = e^{i\nu\pi}\psi_{\text{rel}}(\vec{r}),
\end{equation}
where $\nu$ is the statistical parameter. The case with $\nu = 0$ corresponds to bosonic statistics and that of $\nu = 1$ corresponds to fermionic statistics. For this two-particle situation, projecting to the LLL is similar to the standard case of the single-particle in two-dimensions subject to a large transverse magnetic field \cite{shankarPrinciplesQuantumMechanics1994,jainModelTunnelingProblems1988b}, but with a key difference accounting for fractional statistics. Note that fractional charge $e^*$ can also be accounted for by replacing the electron's charge $e$ with it, where $e^*=\nu e$ for Laughlin quasiparticles. For a description wherein a symmetric gauge is associated with the transverse magnetic field, it is most convenient to employ the complete basis of eigenstates associated with the angular momentum operators. While the COM degree of freedom respects the same form as for the single-particle, $L_{\text{COM}}\ket{l}_{\text{COM}} = l\ket{l}_{\text{COM}}$ the relative coordinate must encode fractional statistics in its boundary condition. Importantly, the states with anyonic statistics have fractional quasi-
angular momenta, $L_{\text{rel}}\ket{k,\nu}_{\text{rel}} = (2k+\nu)\ket{k,\nu}_{\text{rel}}$. The real-space representation of these angular momentum states shows that Eq. (\ref{eq:boundary-conditions}) is explicitly satisfied: $\psi_{k,\nu}(z) = \frac{1}{\sqrt{\pi \Gamma(2k+\nu + 1)}}z^{2k+\nu}e^{-\frac{1}{2}\abs{z}^2}$.\cite{kjonsbergAnyonDescriptionLaughlin1997a}

The two-particle coherent states can be written in the COM and relative coordinates as a product of decoupled coherent states in the COM and relative coordinates\cite{hanssonDimensionalReductionAnyon1992}. The COM coherent state $\ket{Z}$ behaves as a single-particle coherent state and it is the eigenstate of the COM lowering operator $A\ket{Z} = Z$. It is given as a superposition of the COM quasi-angular momentum states:
\begin{equation}
    \ket{Z}  = e^{-\abs{Z}^2/2}\sum_{l=0}^\infty \frac{Z^l}{\sqrt{k!}}\ket{l}_{\text{COM}},
\end{equation}
where $Z = \frac{X-iY}{\sqrt{2}l_B}$ is the complex COM coordinate. One can apply a similar treatment to define the relative coordinate, however, the eigenstates of the linear lowering operator do not satisfy the required anyonic boundary conditions in Eq. (\ref{eq:boundary-conditions}). Thus, the relative coordinate coherent state $\ket{z}_\nu$ is defined to be the eigenstate of the square of the lowering operators $ \frac{a^2}{2}\ket{z}_\nu = \frac{z^2}{2}\ket{z}_\nu$. The state is thus given as a superposition of the relative coordinate quasi-angular momentum states:
\begin{equation}
    \ket{z}_\nu  =N_{z,\nu}\sum_{k=0}^\infty \frac{(z^2/2)^k}{\sqrt{k!\Gamma(k+\nu+1/2)}}\ket{k,\nu}_{\text{rel}},
\end{equation}
where $z = \frac{x-iy}{\sqrt{2}l_B}$ is the complex relative coordinate and $N_{z,\nu}$ is a normalization factor depending on $z$ and $\nu$ \cite{kjonsbergAnyonDescriptionLaughlin1997a}. Thus the full two-particle 
coherent state can be written as a tensor product of the COM and relative coherent states:
\begin{equation}
    \ket{Z,z} = \ket{Z}\otimes\ket{z}_\nu.
\end{equation}
The two-particle coherent states can be characterized using the relative coordinate Husimi-Q quasiprobability distribution given by $f_Q(z,t) = \frac{1}{\pi}\abs{\bra{z}e^{-iHt/\hbar}\ket{z_0}}^2$ for the case in which the COM coordinate at the origin; a generalized version would take into account a more complicated COM dependence.  Figure. \ref{fig:initial_anyon_state} depicts a visualization of this Husimi-Q distribution obtained from our coherent state treatment.

\par \textit{Quadratic potentials and $\mathfrak{su}(1,1)$ algebra.}- Key to our presentation, we characterize the complete set of quadratic operators for the relative coordinates in the context of the LLL for two reasons. First, these operators are generators of the $\mathfrak{su}(1,1)$ algebra associated with transformations that preserve non-commutativity in the LLL (area-preserving deformations in two-dimensions)\cite{haldaneHallViscosityIntrinsic2009,subramanyanCuriousCaseInverted2023a}. These operators together form a mathematical construct that lies at the heart of the geometry of LLL states and that we rely on heavily to offer elegant analytic solutions to the quantities we compute. Second, these operators can together describe the dynamics of particles in the presence of some potential landscapes that are of direct relevance to the key concepts underlying the anyon detection experiments, namely beam-splitter physics and interferometry \cite{fertigTransmissionCoefficientElectron1987,yurkeSU2SU11Interferometers1986,halperinTheoryFabryPerotQuantum2011}.

The generators of the $\mathfrak{su}(1,1)$ algebra in question take the following form: 
\begin{equation}
    B_- = \frac{a^2}{2}\quad B_+ = \frac{(a^\dagger)^2}{2} \quad B_0 = \frac{a^\dagger a+\frac{1}{2}}{2}.
\end{equation}
These operators respect the associated $SU(1,1)$ commutation relations $[B_+,B_-]= -2B_0$  and $[B_0,B_\pm]= \pm B_\pm$. 

The Hamiltonian for all non-interacting quadratic potentials at a given rotational orientation can be expressed generically in terms of these operators as 
\begin{equation}\label{eq:full-hamiltonian}
    \quadh =  \epsilon(B_+ + B_- )  + 2\omega B_0,
\end{equation}
where $\epsilon$ and $\omega$ are the two free parameters that set the energy scale.  Other orientations can be achieved by rotating the coordinate axes. It is assumed that the energy scale of the applied potential is much less than the Landau level splitting such that the system remains in the LLL. 

Thus far, we have discussed operators as applicable to the relative coordinates. The COM coordinates would have an analogous set of operators. As we only consider quadratic operators, the Hamiltonian describing any such potential landscape can be decoupled in terms of the relative and COM coordinates.

 
 The time evolution of the two-particle coherent state can therefore be computed by evolving the COM and relative coordinates separately as follows:
\begin{equation}\label{eq:time-evolution}
    \ket{Z(t),z(t)} \equiv  e^{-i\quadh_{COM} t} \ket{Z} \otimes e^{-i\quadh_{rel}t}\ket{z}_\nu. 
\end{equation}
Thus, via this decoupling, the COM dynamics can be described by that of a single-particle LLL coherent state subject to the appropriate quadratic potential.  The relative coordinate, however has modified static and dynamic features due to the anyonic boundary conditions and it contains all the information on fractional statistics.  


We now employ this framework to analyze the dynamics in the presence of unbounded and bounded potentials, respectively. The closed-form BCH relations of the $\mathfrak{su}(1,1)$ algebra \cite{martinez-tibaduizaNewBCHlikeRelations2020} allow us to compute unitary time evolution described by Eq. (\ref{eq:time-evolution}), and hence compute the dynamic trajectories of anyons in these landscapes.



\par
\begin{figure}[h!]
  \centering
  \subfigure[]{%
    \includegraphics[width=0.48\linewidth]{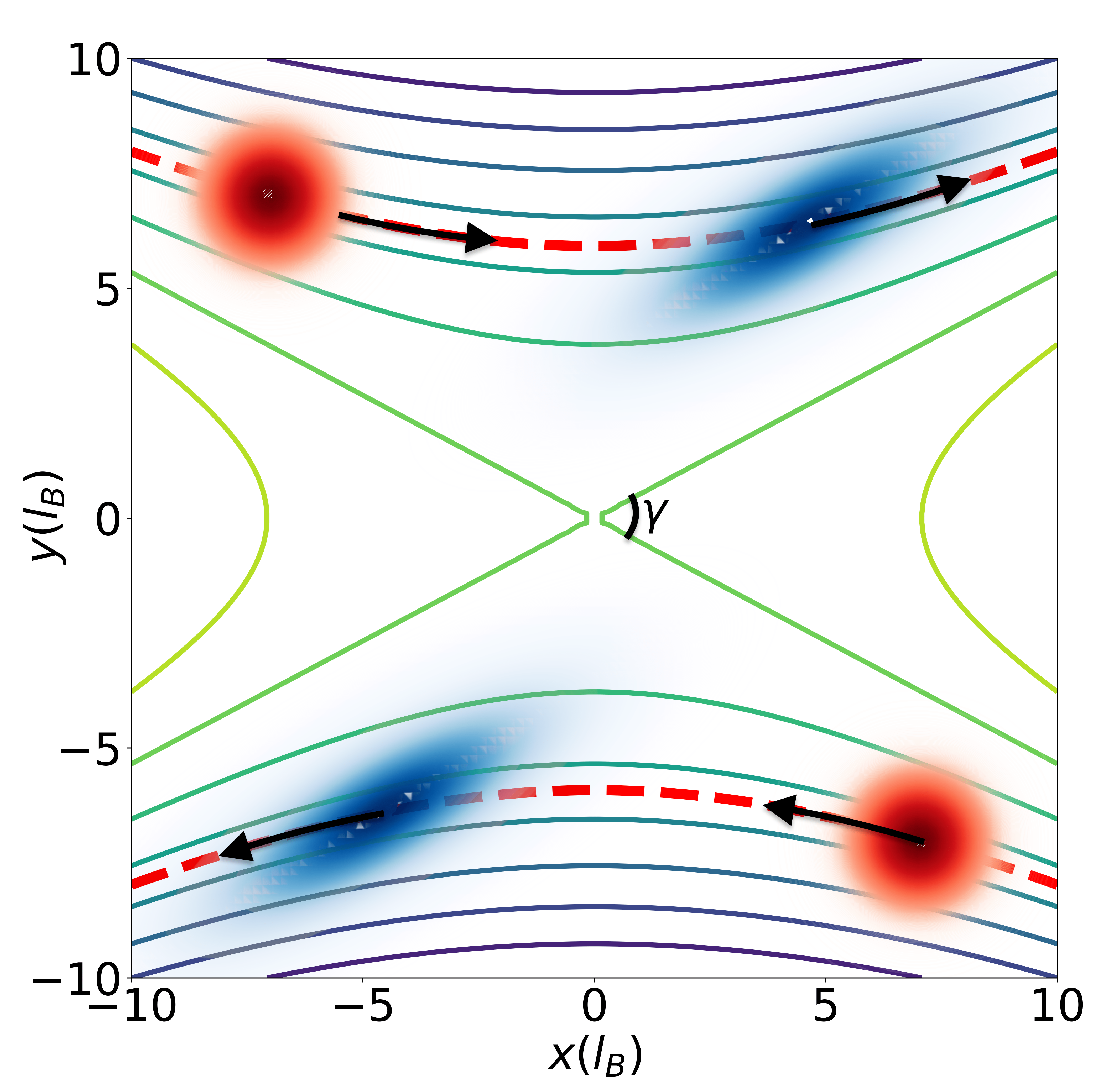}
    \label{fig:saddle_trajectory}%
  }
  \hfill
  \subfigure[]{%
    \includegraphics[width=0.48\linewidth]{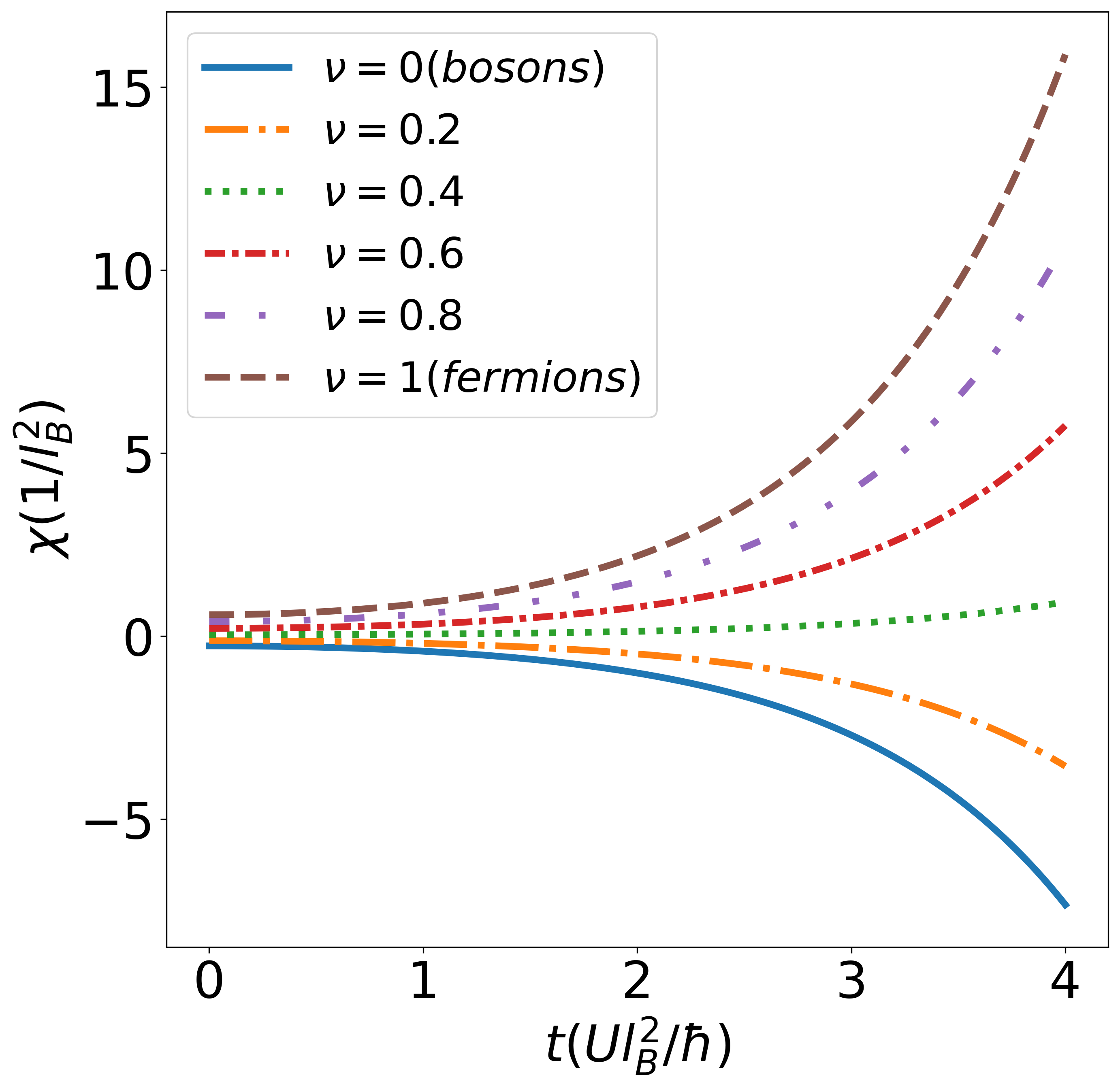}
    \label{fig:saddle_bunching}%
  }
  \caption{(a) Dynamics of two anyons under a saddle potential, shown overlaid on equipotential lines at two different times. Squeezing as the anyon centers follow equipotential contours is visualized.  
  (b) Time evolution of the bunching parameter under saddle potentials for various values of $\nu$ with an initial seperation of $l_B$. While fermions always separate along opposite legs of the saddle and bosons bunch along the same one, anyons can do either depending on the initial condition.}
\end{figure}

 \textit{Unbounded potentials.}-
The general quadratic form described by Eq. (\ref{eq:full-hamiltonian}) allows for unbounded dynamics for parameters that correspond to  saddle potentials\cite{matthewsScatteringTheoryQuantum2009}, such as shown in Fig. \ref{fig:saddle_trajectory}. Such potentials can act as beam splitters wherein incoming quasiparticles can split off into one of two directions\cite{vishveshwaraCorrelationsBeamSplitters2010,subramanyanCorrelationsDynamicsInterferometry2019}. Key experiments detecting fractional statistics via current correlation measurements have relied on such beam splitter concepts. While they have employed edge-state quasiparticles, scattering happens in the bulk through point contact geometries; here, the bulk is in fact subject to a saddle potential. Moreover, saddle potentials are known to occur ubiquitously in solid state QH landscapes where disorder creates valleys and hills, and saddles in between \cite{huckesteinScalingTheoryInteger1995}.

In terms of spatial coordinates, saddle potentials can be described as
\begin{equation}
     \quadh = Ux^2 - Vy^2, \label{eq:saddle-hamiltonian}
\end{equation}
 where $U>0$ and $V>0$. As shown in Fig. \ref{fig:saddle_trajectory}, the angle made by the principal axes of the saddle potential is given by $\gamma = 2\arctan{\left(\sqrt{\frac{U}{V}}\right)}$. The most general form for a saddle potential would involve a rotation of the axes of symmetry by an angle $\theta$. Dynamics under saddle evolution is described by Eq. (\ref{eq:time-evolution}) with the identification $\epsilon = \frac{l_B^2}{2}(U+V)$ and $\omega = \frac{l_B^2}{2}(U-V)$, therefore respecting the inequality $\abs{\epsilon}>\abs{\omega}$. As was shown in previous work, including by two of the current authors \cite{subramanyanCorrelationsDynamicsInterferometry2019,vishveshwaraCorrelationsBeamSplitters2010}, a close analogy with quantum optics reveals that the symmetric saddle potential ($U = V$) in the LLL plays the role of a `squeezing operator`. Squeezing operators are the operators that generate Bogoliubov transformations, and can be generically defined as
\begin{equation}
    \hat{S}(t) = e^{\frac{1}{2}\left(\xi A^2 - \xi^* (A^\dagger )^2\right)},
\end{equation}
where for a symmetric saddle $\xi = \frac{il_B^2 U t}{\hbar}$. The magnitude of $\xi$ is known as the `squeeze parameter` $\beta = \abs{\xi}$. Half the argument of $\xi$ is the `squeeze angle` $\phi = \frac{1}{2}\arg(\xi)$: the angle along which squeezing occurs compared to the $x$-axis. In the beam splitter model, the angle $\phi$ determines the orientation along which the uncertainty in phase difference between the two beams is minimal \cite{yurkeSU2SU11Interferometers1986}.

At a formal level, we first focus on single-particle coherent state dynamics by considering the COM time-evolution $\ket{Z(t)}$ in Eq. (\ref{eq:time-evolution}). In the analogy with optics, the non-commuting nature of spatial coordinates $(X,Y)$ in the LLL translates to that of phase space operators of position and conjugate momentum. Thus, first off, because the semi-classical velocity of the particle can be found to be orthogonal to the gradient of the potential 
\cite{shankarPrinciplesQuantumMechanics1994}, the guiding center coordinates time-evolve by traveling along an equipotential line of the saddle that is determined by its initial placement. 

In this quadratic case, we can use BCH relations \cite{martinez-tibaduizaNewBCHlikeRelations2020} to precisely show the time evolution for an asymmertic saddle ($U \ne V$) is the same as first rotating the coherent state about the origin in the $X-Y$ plane by a time dependent angle  $\delta(t) = \frac{\pi}{2}+\tan^{-1} \left( \frac{2\sqrt{UV}}{U-V}\coth(\frac{t}{t^{\prime} }) \right)$, where $t^{\prime}  =\frac{\hbar}{l_B^2\sqrt{UV}}$, and then applying a squeezing operator with a time dependent squeezing parameter $\beta(t)  = \cosh^{-1} \left(\sqrt{\cosh^2\left(\frac{t}{t^{\prime} }\right) + \frac{(U-V)^2}{4UV}\sinh^2\left(\frac{t}{t^{\prime} }\right)} \right)$ and a time dependent squeeze angle that is related to the rotation angle $\phi(t) = \frac{\delta(t)}{2}-\frac{\pi}{4}$. These parameters can be used the compute the mean COM coordinate as a function of time: $\langle Z(t)\rangle=i^{i\phi(t)}\cosh(\beta(t))Z_0-ie^{-2i\phi(t)}\sinh(\beta(t))Z_0^*$. From the explicit form it can be shown that the COM follows the equipotential line of the asymmetric saddle potential as expected.

Turning to two-particle physics, we make an electronic analogy to the photon bunching seen in two-photon correlations between two coherent beams of light, as first described by Hanbury Brown and Twiss\cite{brownCorrelationPhotonsTwo1956}. The bunching exhibited by light is quantum mechanically a result of the bosonic nature of photons. Since bosons tend to bunch and fermions tend to anti-bunch, we can introduce a bunching parameter that measures the "attraction/repulsion" experienced due to quantum statistics in comparison to distinguishable particles \cite{townsendModernApproachQuantum2012}. As in previous work \cite{subramanyanCorrelationsDynamicsInterferometry2019}, we employ the following two-particle correlator as a definition of the bunching parameter: 
\begin{equation}
    \chi_\nu(t) \equiv\frac{1}{4l_b^2} \left[\bra{z(t),\nu}\hat{r}^2\ket{z(t),\nu}-\bra{z(t)}_d\hat{r}^2\ket{z(t)}_d\right],
\end{equation}
where $\bra{z(t)}_d\hat{r}^2\ket{z(t)}_d$ refers to the expectation of the separation for two distinguishable particles in the same potential. The bunching parameter for anyonic coherent states with no time evolution was computed in  Ref. \cite{subramanyanCorrelationsDynamicsInterferometry2019}. As expected, bosons bunch ($\chi < 0$) and fermions anti-bunch ($\chi > 0$). Anyons tend to display properties of either bosons or fermions, as determined by the sign of $\chi$,  depending on their separation. 
The temporal behavior of $\chi$ resulting from the application of the saddle potential is shown in  Fig. \ref{fig:saddle_trajectory}. For instance when $Z = 0$, with close enough initial separation $z_0$, the two particles begin on diametrically opposite sides of the origin. The peaks of the coherent state then follow along their respective equipotential lines until their closest approach when they may partially overlap. After this point, the anyons continue along their respective equipotential lines, being squeezed in a qualitatively similar manner to a single particle. Depending on $\nu$, they either follow a boson-like path corresponding to a $\chi<0$ (bunching), or a fermion-like path corresponding to $\chi>0$. The initial separation and statistical parameter $\nu$ of the anyons determines the sign of $\chi$. Its magnitude grows exponentially with time as dictated by the saddle potential evolution, making quantum statistical properties more pronounced. This is seen in Fig. \ref{fig:saddle_bunching}, where the bunching parameter is plotted as a function of time for different values of $\nu$ for a given initial separation. There is a clear separation in bunching parameters based on $\nu$.

\par \textit{Bounded potentials.}-
While unbounded and bounded potentials host vastly different behavior, formally, we may replace the saddle form in Eq. (\ref{eq:saddle-hamiltonian}) to the elliptical form
\begin{equation}
     \quadh = Ux^2 + Vy^2 \label{eq:elliptical-hamiltonian},
\end{equation}
once more with $U>0$ and $V>0$. This describes an ellipse entered at the origin with eccentricity $e =\sqrt{1 - \frac{U}{V}}$. Rotated elliptical potentials can be achieved by the same coordinate transformation that gave rise to rotated saddle potentials discussed above. However, the identifications of $\epsilon$ and $\omega$ as defined after Eq. (\ref{eq:saddle-hamiltonian}) in terms of $U$ and $V$ are interchanged with $\epsilon = \frac{l_B^2}{2}(U-V)$ and $\omega = \frac{l_B^2}{2}(U+V)$, and crucially, $\abs{\epsilon}<|\omega|$. Effectively, compared with the saddle potential, this inequality renders $t^{\prime}$ to be imaginary. Therefore, the hyperbolic forms associated with Bogoliubov transformations in the squeeze parameter $\beta(t)$ and rotation angle $\delta(t)$ convert to oscillatory sinusoidal forms.

As with the unbounded potentials, as per expectation, each of the coherent states traverses along equipotential lines. Once again, formally, we can first focus on single-particle dynamics by analyzing the COM motion $\ket{Z(t)}$. In this case of bounded potentials, the oscillatory sinusoidal forms therefore correspond to guiding center time-evolution along ellipsoidal equipotential lines for the same reason as in the saddle. We can write down the rotation angle and squeeze parameter as before, but now with the identification $\delta(t) =\frac{\pi}{2}+\tan^{-1} \left( \frac{2\sqrt{UV}}{U+V}\cot(\frac{t}{t^{\prime} }) \right)$ and $\beta(t) = \cosh^{-1} \left(\sqrt{\cos^2\left(\frac{t}{t^{\prime} }\right) + \frac{(U+V)^2}{4UV}\sin^2\left(\frac{t}{t^{\prime} }\right)} \right)$ and a squeeze angle $\phi(t) = \frac{\delta(t)}{2}-\frac{\pi}{4}$. Where $\frac{1}{t^\prime}=\frac{l_B^2\sqrt{UV}}{\hbar}$ now corresponds to the angular frequency of the oscillations. Because of the oscillatory nature of $\beta(t)$, the ellipse causes the COM states to oscillate between an unsqueezed state and a maximally squeezed state with the same angular frequency $\frac{1}{t^{\prime} }$ as that of the motion of the COM state. The squeezing becomes more pronounced as the ellipse gets more eccentric. 

Turning now to two-particle dynamics, we can show that for the symmetric case where $U = V$, the relative coordinate coherent state evolves as 
\begin{equation}
    \ket{z(t)}_\nu = e^{iUtl_B^2(\nu/2+1/2)/\hbar}\ket{z_0e^{iUtl_B^2/{2\hbar}}}_\nu.
\end{equation}
 Thus, "elementary exchange"\cite{readPerspectiveAnyonicBraiding2023} corresponds to $Utl_B^2/\hbar = 2\pi$. At this time a phase of $e^{i\pi\nu}$ is picked up by the state, satisfying the boundary conditions. This is demonstrated in Fig \ref{fig:worldlines} where the worldlines of the two particles can be seen braiding \cite{parisBraidGroupsArtin2009}. Since the statistical phase is topologically invariant in the path taken by the anyons, and we can smoothly deform from the circle($U=V$, $U,V>0$) to the ellipse ($U \ne V$, $U,V>0$), we can conclude that the same phase is picked up by an elementary exchange under the influence of an asymmetric, elliptical potential. Similar to the saddle potential we can first look at a case where $Z = 0$. Here, the two particles orbit their center of mass, remaining on opposite sides of the origin. In this case a half orbit corresponds to an elementary exchange of the two particles, which corresponds to a phase of $e^{i\pi\nu}$ being picked up by the state. We can also look at the case where one of the particles is at the origin and the other is not. In this case, the particle not at the origin will orbit the particle at the origin. Here a half oscillation does not correspond to an exchange, while a full oscillation corresponds to a double exchange of the two particles, which picks up a phase of $e^{2i\pi\nu}$. This second corresponds to the interferometer setting in which one quasiparticle encircle another one trapped on an island of quantum Hall fluid; our treatment offers an exact description for two anyons. 

\begin{figure}[h!]
    \centering
    \subfigure[]{%
        \includegraphics[width=0.48\linewidth]{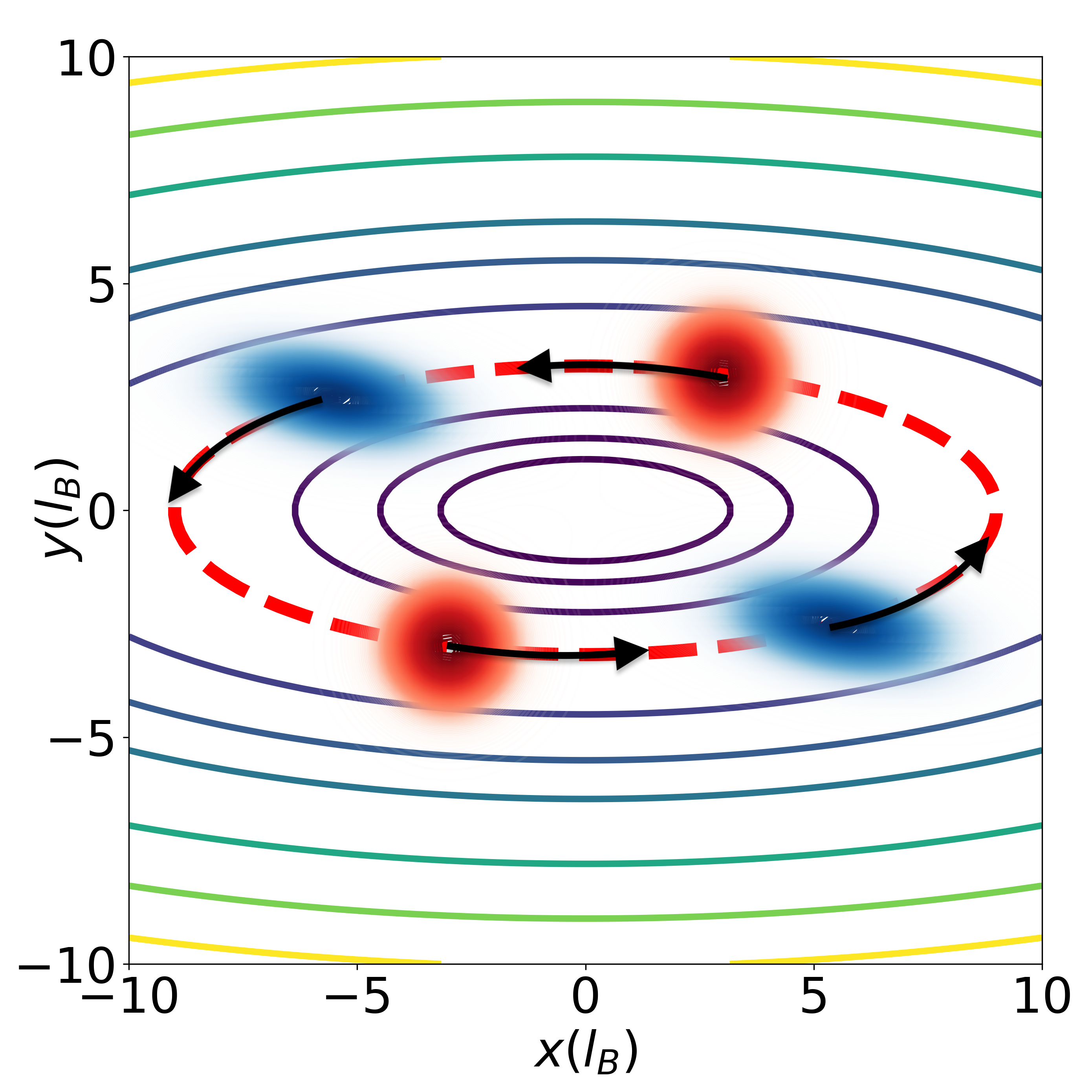}
        \label{fig:ellipse_trajectory}%
    }
    \hfill
    \subfigure[]{%
        \includegraphics[width=0.48\linewidth]{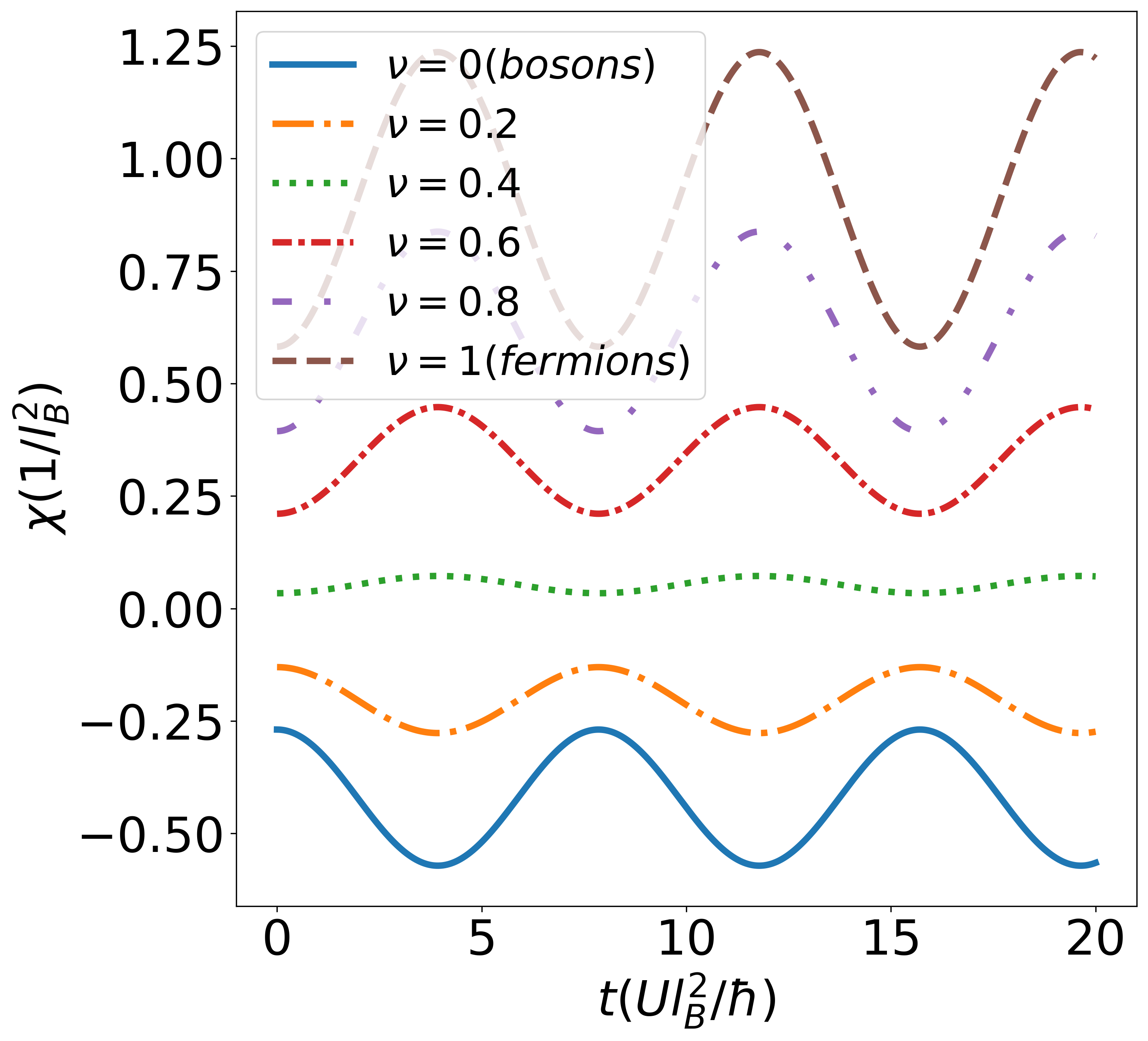}
        \label{fig:ellipse_bunching}%
    }
    \caption{(a) Dynamics of two anyons under an elliptical potential, shown overlaid on equipotential lines at two different times. Squeezing behavior as the anyon centers evolve along equipotential contours is illustrated.  
    (b) Time evolution of the bunching parameter under elliptical potentials for various values of $\nu$ with an initial separation of $l_B$. The amplitude and mean of the oscillations varies depending on $\nu$}
    \label{fig:ellipse_trajectory_and_bunching}
\end{figure}

The trajectory of anyons under the influence of an elliptical potential is shown in Fig. \ref{fig:ellipse_trajectory} for $Z=0$. It can be seen that the anyons are periodically squeezed and un-squeezed with a maximum amplitude of the squeezing parameter growing with the eccentricity of the potential, with the single-particle case. The bunching parameter also oscillates as seen in Fig. \ref{fig:ellipse_bunching}. As the eccentricity of the ellipse increases, the amplitude of oscillations in the bunching parameter become larger. For the circle, the bunching parameter remains constant. The statistical parameter $\nu$ and initial separation also affect the sign and amplitude of oscillations, starting with positive amplitudes for fermions($\nu = 1$) and smoothly transitioning to negative amplitude for bosons ($\nu = 0$). This is consistent with the fact that bosons tend to bunch and fermions tend to anti-bunch. Similar to the saddle case, there is a clear seperation in the time-averaged bunching parameter based on $\nu$.\newline

\par \textit{Conclusions}-
In conclusion, we have presented a robust framework to describe key features of localized FQH bulk anyons and their dynamics in the generic quadratic potentials. We build on the coherent state description for the LLL projection of two quasi-hole Laughlin state with anyon statistical parameter $\nu$. Parallels with quantum optics combined with the $\mathfrak{su}(1,1)$ algebra strutcure underlying the coherent states enables us to perform an in-depth treatment. Unbounded saddle potentials capture the essential features of beam-splitter/collider and Hanbury Brown-Twiss physics while bounded elliptical potentials capture motion in trapped geometries and the explicit phase evolution relevant to interferometry. Comparison with other approaches, such as semiclassical treatments \cite{subramanyanDynamicsClassicalBosons2024,hanssonClassicalPhaseSpace2001}, are in order. Two immediate and outstanding directions to extend this framework entail connecting edge state studies with these bulk analyses and generalizing to the $\nu = 5/2$ state, the frontrunner for non-Abelian anyons. Experimentally, beyond traditional solid-state quantum Hall platforms, realization of FQH states in topological insulators, ultracold atoms or metamaterials might offer fertile ground for patterning, nucleating and manipulating bulk anyons in the theoretical situations envisioned here.

\par 
We gratefully acknowledge that this work has been supported by the
National Science Foundation through Grant No. DMR-
2004825. We thank H. Hansson, B. Rhyno, and A. Yazdani for the many insightful discussions. 
\bibliography{PRL_symmetries_dynamics_qhbulk.bib}
\bibliographystyle{unsrt}
\end{document}